\documentclass[aps,nofootinbib,onecolumn,citeautoscript]{revtex4-2}
\usepackage{amsmath,amssymb}
\usepackage{enumerate}
\usepackage{xcolor}
\usepackage[colorlinks=true,citecolor=violet,linkcolor=red,linkbordercolor=cyan]{hyperref}
\usepackage{setspace}
\usepackage{enumerate}
\usepackage{graphicx}
\usepackage{float}

\begin{document}
\title{Comparative Dynamical Study of a Bound Entangled State}
\author{Suprabhat Sinha$^\ast$\footnote[0]
{$^\ast$suprabhatsinha64@gmail.com} \\\vspace{0.3cm} \textit{School of Computer Science, Engineering and Applications, \\ D Y Patil International University, Akurdi, Pune-411044, India}}

\begin{abstract}

The bound entangled state carries noisy entanglement and it is very hard to distill but the usefulness of bound entangled states has been depicted in different applications. This article represents a comparative dynamical study of an open quantum system for one of the bound entangled states proposed by Bennett \textit{et al.} The study is conducted under the influence of Heisenberg, bi-linear bi-quadratic and Dzyaloshinskii–Moriya (DM) interaction. During the study, an auxiliary qutrit interacts with one of the qutrits of the selected two qutrit bound entangled state through different interactions. The computable cross-norm or realignment (CCNR) criterion has been used to detect the bound entanglement of the state and the negativity has been applied to measure the free entanglement. From this three-fold study it is observed that, although the auxiliary qutrit plays a significant role during the interaction, the probability amplitude of the qutrit does not affect the open quantum system. Further, it is found that the Dzyaloshinskii–Moriya (DM) interaction performs better to activate the chosen bound entangled state among all the interactions.

\end{abstract}

\maketitle

\section{Introduction}
Quantum entanglement is one of the cornerstones, which plays a major role in developing the applications of quantum computation and quantum information theory \cite{NL-CH, QE, NG}. In most of the applications of quantum computation and quantum information processing, maximally entangled quantum states are preferred for perfect execution. The quantum entangled states can be divided into two types; free entangled quantum states and bound entangled quantum states \cite{HO1}. Free entangled quantum states are distillable states and pure entanglement can be turned out easily. Due to this reason, these states can be effectively utilized for quantum computation and information processing. A wide range of applications of the free entangled quantum states have been investigated in quantum teleportation \cite{QT1, QT2, QT3}, quantum cryptography \cite{QC1, QC2}, quantum sensing \cite{QS}, quantum imaging\cite{QI}, quantum games \cite{QG} and so many domains. On the other side, bound entangled quantum states are noisy entangled quantum states. This type of quantum state is very hard to distill and extract pure entanglement. For this reason, bound entangled quantum states are generally avoided in quantum computation and information processing. In some of the recent studies, it is found that the bound entangled quantum states may be useful for distributing quantum keys securely \cite{KEY}, hiding quantum data \cite{DATA}, concentrating remote quantum information \cite{RE}, reducing communication complexity \cite{COM} and so on. It has also been discussed that with some free entanglement, a bound entangled state enhances the teleportation fidelity \cite{QT4, QT5}.

The term `Bound Entanglement' was first introduced by Horodecki \textit{et al.} and the preliminary bound entangled quantum state is also proposed by them \cite{HO2}. After that many people, Bennett \textit{et al.} \cite{BT1}, Jurkowski \textit{et al.} \cite{JKI1} etc. have contributed to construct different bound entangled quantum states. Recently, experimental manifestation and distillation of the bound entanglement have been realized \cite{E1, E2, E3, E4}. From the application point of view, the evolutionary dynamics and the distillation of the bound entangled quantum states are as important as the construction. Different authors propose different methods for the dynamical analysis and distillation of different bound entangled quantum states. Guo-Qiang \textit{et al.} have shown the time evolution of a bound entangled quantum state provided by Horodecki \textit{et al.}, under bi-linear bi-quadratic interaction \cite{BI} and offers a way to free bound entanglement. Baghbanzadeh \textit{et al.} have presented the distillation of three bound entangled quantum states proposed by Horodecki \textit{et al.} and Bennett \textit{et al.} using weak measurements \cite{WK}. Sharma \textit{et al.} have also studied one of the Horodecki \textit{et al.} bound entangled quantum states. But they use Dzyaloshinskii–Moriya (DM) interaction to show time evolution and distillation \cite{K1}. Recently, they have made their work one step forward and repeat their study for a bound entangled quantum state provided by Jurkowski \textit{et al.} \cite{K2}. During these studies, different tools are used to detect and measure entanglement. At present, there are several mathematical tools available for characterization, detection, and measurement of the free entanglement for bipartite quantum systems \cite{EM}. Negativity is one of the measures among them, which is also a good tool for the quantification of the free entanglement \cite{N}. On the other hand, the characterization and detection of the bound entanglement is still an open problem. Although some criteria have been already developed to detect the bound entanglement, such as separability criterion, realignment criterion, computable cross-norm or realignment (CCNR) criterion \cite{C1, C2, C3, C4, C5} etc.

In the current study, comparative dynamical analysis and distillation of a $3\times3$ dimensional bipartite bound entangled quantum state provided by Bennett \textit{et al.}, is investigated. The analysis is explored with the help of an auxiliary qutrit under three different physical interactions; Heisenberg \cite{H1,H11,H12}, bi-linear bi-quadratic \cite{H2,H21} and DM interaction \cite{DM1,DM2,DM3}. The selected interactions have already shown their efficacy in the field of quantum computation and information \cite{K3, K4, K5, K6, K7, K8, K9, K10, K11, K12, K13}.  According to the best of my knowledge, this type of study under the chosen interactions with the considered bound entangled state is missing in the literature. During the study CCNR criterion has been selected to detect the bound entanglement of the system and negativity has been applied for quantifying and measuring the free entanglement of the system.

The study is sketched as follows. The discussion about the selected bound entangled state, negativity, and CCNR criterion is done in section 2. Section 3 deals with the unitary dynamics and the Hamiltonian of the different interactions in the open quantum system. In section 4, the dynamical behavior for different interactions has been studied. In the last section, the conclusion of the results is explained.

\section{Bound entangled state, Negativity and CCNR criterion}
In the current section, the bound entangled state proposed by Bennett \textit{et al.}\cite{BT1}, negativity, and CCNR criterion are discussed. The state is a $3\times3$ dimensional bipartite state dealing with two qutrits $A$ and $B$. The density matrix of the specified bound entangled state can be written as,

\begin{equation}
\rho_{AB}=\frac{1}{4} \left( (I \otimes I)-\sum_{i=0}^4\vert \psi_i\rangle \langle\psi_i \vert \right).
\label{bes}
\end{equation} 
Where, $I$ is the $3\times3$ dimensional identity matrix,
\begin{eqnarray*}
\vert\psi_0\rangle=\frac{1}{\sqrt{2}} \vert 0\rangle (\vert 0\rangle-\vert1\rangle),\quad
\vert\psi_1\rangle=\frac{1}{\sqrt{2}} (\vert0\rangle-\vert1\rangle)\vert2\rangle,\quad
\vert\psi_2\rangle=\frac{1}{\sqrt{2}}\vert2\rangle \vert1\rangle-\vert2\rangle),
\end{eqnarray*}
\begin{eqnarray*}
 \vert\psi_3 \rangle=\frac{1}{\sqrt{2}}(\vert1\rangle-\vert2\rangle)\vert0\rangle \quad \text{and} \quad
\vert\psi_4\rangle=\frac{1}{3} (\vert0\rangle+\vert1\rangle+\vert2\rangle)(\vert0\rangle+\vert1\rangle+\vert2\rangle).
\end{eqnarray*}

Negativity and CCNR criterion are primarily used to detect and quantify the entanglement in our current work. The negativity has been used to quantify the free entanglement while the CCNR criterion has been used to detect the bound entanglement of the system. CCNR is a very simple and strong criterion for the separability of a density matrix. This criterion can detect a wide range of bound entangled states and performs with better efficacy. The negativity $(N)$ and CCNR criterion are defined as below,
\begin{equation}
N=\frac{(\left \|\rho_{AB}^{T}\right \|-1)}{2} \label{N}
\end{equation} 
and
\begin{equation}
CCNR=\left\|(\rho_{AB}-\rho_{A}\otimes \rho_{B})^R\right\|-\sqrt{(1-Tr \rho_{A}^{2}) (1-Tr \rho_{B}^{2})}. \label{CCNR}
\end{equation}
Where $\|..\|$, $(..)^T$ and $(..)^R$ represent the trace norm, partial transpose and realignment matrix respectively. Further $\rho_{A}$ and  $\rho_{B}$ are the reduced density matrices of qutrit A and qutrit B and $\rho_{AB}$ is the density matrix of the bound entangled state respectively and expressed as,
\begin{eqnarray*}
\rho_{A}=Tr_{BC}(\rho_{ABC}), \quad \rho_{B}=Tr_{AC}(\rho_{ABC})\quad \text{and} \quad \rho_{AB}=Tr_{C}(\rho_{ABC}).
\end{eqnarray*}

For a system, $N>0$ or $CCNR>0$ implies that the state is entangled, $N=0$ and $CCNR>0$ implies that the state is bound entangled and $N>0$ corresponds to the free entangled state.

\section{Unitary dynamics and Hamiltonian of different interactions}
In this section, the interaction between the closed system and the auxiliary qutrit is presented. Further, the unitary dynamics of the system and the Hamiltonian of the different interactions are also discussed. During the present study, it is considered that the auxiliary qutrit $(C)$ interacts with one of the qutrits of the pair in the closed system through different interactions (Heisenberg, bi-linear bi-quadratic, and DM). Here the closed system is the bound entangled state provided by Bennett \textit{et al.} and consists of two qutrits ($A$ and $B$). In the current article, it is assumed that the interaction takes place only in the Z-direction between qutrit $A$ and auxiliary qutrit $C$. The state vector of the additional auxiliary qutrit $(C)$ can be expressed as,
\begin{equation}
\vert C\rangle=\alpha\vert0\rangle+\beta\vert1\rangle+\gamma\vert2\rangle\label{eqt}
\end{equation}
with the normalization condition,
\begin{equation}
\vert\alpha\vert^{2}+\vert\beta\vert^{2}+\vert\gamma\vert^{2}=1.\label{nc}
\end{equation}
The density matrix of the qutrit $C$ reads as,
\begin{equation}
\rho_{C}=\left[ \begin{array}{ccc}
\vert\alpha\vert^2&\alpha\beta&\alpha\gamma \\
\alpha\beta&\vert\beta\vert^2&\beta\gamma \\
\alpha\gamma&\beta\gamma&\vert\gamma\vert^2 \\
\end{array}
\right ]. \ \ \ \label{edm'}
\end{equation} 
Using the Eq.\ref{nc} the above density matrix can be written as, 
\begin{equation}
\rho_{C}=\left[ \begin{array}{ccc}
\vert\alpha\vert^2&\alpha\beta&\alpha\sqrt{1-\vert\alpha\vert^2-\vert\beta\vert^2} \\
\alpha\beta&\vert\beta\vert^2&\beta\sqrt{1-\vert\alpha\vert^2-\vert\beta\vert^2}  \\
\alpha\sqrt{1-\vert\alpha\vert^2-\vert\beta\vert^2}  &\beta\sqrt{1-\vert\alpha\vert^2-\vert\beta\vert^2} & 1-\vert\alpha\vert^2-\vert\beta\vert^2 \\
\end{array}
\right ]. \ \ \ \label{edm}
\end{equation}
After interaction, the initial density matrix of the open system can be expressed as below,
\begin{equation}
\rho_{ABC}(0)=\rho_{AB} \otimes \rho_{C}.  \label{sdm}
\end{equation}
Since it is assumed that in the current article the auxiliary qutrit $(C)$ interacts with the qutrit $A$ of the bound entangled closed system. So, the Hamiltonian of the interacted system can be written as,
\begin{equation}
H=H_{AB}+H_{AC}^{int}. \label{H'}
\end{equation}
Where $H_{AB}$ is the Hamiltonian of qutrit $A$ and qutrit $B$ and $H_{AC}^{int}$ is the interaction Hamiltonian of qutrit $A$ and qutrit $C$. Here it is considered that qutrit $A$ and qutrit $B$ are uncoupled initially, so $H_{AB}$ is zero. Now the Hamiltonian of the open quantum system becomes,
\begin{equation}
H=H_{AC}^{int}. \label{H''}
\end{equation}

According to the postulates of quantum mechanics, the unitary time evolution of a physical system is obtained from the time-dependent Schr\"odinger equation given below,
\begin{equation}
i\hbar\frac{d}{dt}\vert\psi(t)\rangle=E\vert\psi(t)\rangle.
\end{equation}
Where $E$ is the real energies of the physical system. The solution of the above equation is expressed as,
\begin{equation}
\vert\psi(t)\rangle=e^{\frac{-iHt}{\hbar}}\vert\psi(0)\rangle.\label{se}
\end{equation}
For the application of density matrix, Eq.\ref{se} can be framed as,
\begin{equation}
\rho(t)=U(t) \cdot \rho(0) \cdot U(t)^{\dagger}\label{tm1}.
\end{equation}
Where $U(t)=e^{\frac{-iHt}{\hbar}}$ is the unitary matrix, known as `Time Evolution Operator', includes the Hamiltonian $(H)$ in exponential. To simplify the present study $\hbar$ is assumed as 1 (i.e. $\hbar=1$) and using the Eqs.\ref{sdm} and \ref{tm1} time evolution density matrix of the open system can be written as,
\begin{equation}
\rho_{ABC}(t)=U(t) \cdot \rho_{ABC}(0) \cdot U(t)^{\dagger}\label{tsdm}.
\end{equation}
This time evolution density matrix is used to explain the dynamics of the open system with different interactions in the next section.

Heisenberg interaction is one of the commonly considered interactions in the scientific community. The entanglement dynamics of several quantum states have been explored under this interaction. The Hamiltonian of the open system under this interaction is considered to be XXX model in one dimension without any external magnetic field and following no boundary conditions. Under the above consideration, the Heisenberg Hamiltonian of a quantum system can be expressed as,
\begin{equation}
H_{1}=-J \cdot \sum\limits_{j=1}^N\sigma_j \cdot \sigma_{j+1}\label{h1'}.
\end{equation} 
As per the previous assumption, the interaction happens in the Z-direction between qutrit $A$ and auxiliary qutrit $C$. So, under this assumption, the Heisenberg Hamiltonian of Eq.$\ref{h1'}$ can be framed as,
\begin{equation}
H_{1}=-J \cdot (\sigma_A^z \otimes \sigma_C^z)\label{h1}.
\end{equation} 
Where $J$ is the coupling constant and $\sigma_A^z$ and $\sigma_C^z$ are the Gell-Mann matrices of qutrit A and qutrit C respectively.

Bi-linear bi-quadratic interaction is the non-linear extension of Heisenberg interaction. Under the same consideration of the Heisenberg Hamiltonian, the bi-linear bi-quadratic Hamiltonian of a quantum system can be written as,
\begin{equation}
H_{2}=-J \cdot \sum\limits_{j=1}^N\left[\left(\sigma_j \cdot \sigma_{j+1}\right)+\left(\sigma_j\cdot \sigma_{j+1}\right)^2\right]\label{h2'}.
\end{equation}
According to the interaction assumption of the study, this interaction also occurs in the Z-direction between qutrit $A$ and auxiliary qutrit $C$ and the bi-linear bi-quadratic Hamiltonian of Eq.$\ref{h2'}$ can be rewritten as,
\begin{equation}
H_{2}=-J \cdot \left[\left(\sigma_A^z\otimes \sigma_C^z\right)+\left(\sigma_A^z\otimes \sigma_C^z\right)^2\right]\label{h2}.
\end{equation}
Here $J$ is the coupling constant and $\sigma_A^z$ and $\sigma_C^z$ are the Gell-Mann matrices of qutrit A and qutrit C respectively similar to the Heisenberg Hamiltonian. It is also noted that $\left(\sigma_A^z\otimes \sigma_C^z\right)^2$ is the non-linear term arrived in the bi-linear bi-quadratic Hamiltonian as an extension over Heisenberg Hamiltonian.

DM interaction plays an important role in quantum computation and quantum information. This interaction has some lucid properties and most of the time it enhances the entanglement in a physical system. As the requirement of prolonged entanglement is necessary to execute quantum applications, DM interaction has its gravity in the quantum information community. The DM interaction Hamiltonian of any quantum system can be written as,
\begin{equation}
H_{3}=\vec{D} \cdot (\vec{\sigma_1} \times \vec{\sigma_2}). \label{h3'}
\end{equation}
Where $\vec{D}$ is a vector and its component represents the interaction strength along the direction of the interaction and $\vec{\sigma_1}$ and $\vec{\sigma_2}$ denote the Pauli vectors. As per the previous assumption regarding the interaction between qutrit A and auxiliary qutrit C, the DM interaction Hamiltonian of Eq.$\ref{h3'}$ can be expressed as,
\begin{equation}
H_{3}=D \cdot (\sigma_A^x \otimes \sigma_C^y - \sigma_A^y \otimes \sigma_C^x). \label{h3} 
\end{equation}
Where $D$ is the interaction strength along Z-direction and $\sigma_A^x$, $\sigma_A^y$ and $\sigma_C^x$, $\sigma_C^y$ are the Gell-Mann matrices of qutrit $A$ and qutrit $C$ respectively.

\section{Dynamics of open quantum system}
In this section, the dynamics of the open quantum system are explored using the time evolution density matrix of the system, given in Eq.\ref{tsdm}. This dynamical analysis used negativity and CCNR criteria for detection and quantification of entanglement as previously discussed. The open quantum system is formed through the interaction between qutrit $A$ of the closed system and auxiliary qutrit $C$, as mentioned before. In the current article, three different interactions are considered between qutrits, and discussed the results in three consecutive cases, which are given in the successive subsections. 

\begin{figure*}
\centering
\includegraphics[scale=1.25]{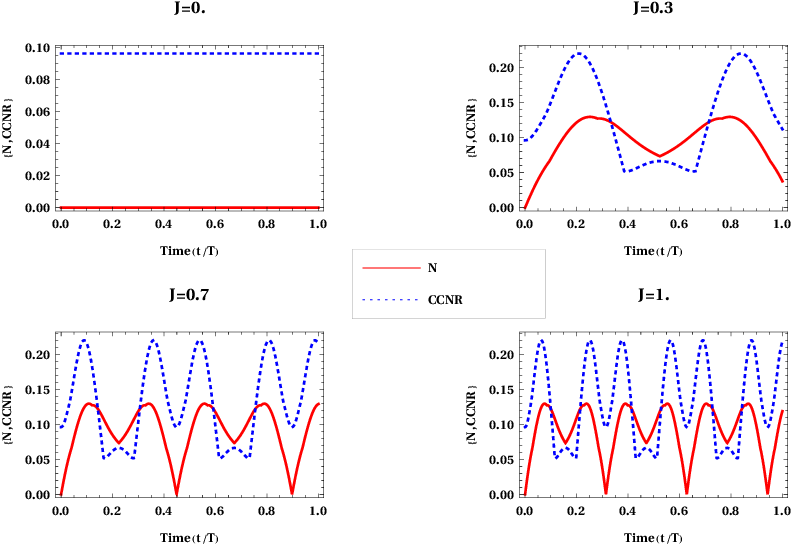} 
\caption{Plot of Negativity(N) and CCNR vs. Time$(t/T)$}\label{f1}
\end{figure*}

\subsection*{Case 1: Dynamics under Heisenberg interaction}  
In this case, it is considered that qutrit $A$ of the closed system and auxiliary qutrit $C$ interact through Heisenberg interaction. The open quantum system dynamics of the bound entangled state is examined under the Heisenberg Hamiltonian for different coupling constant $(J)$ in the range $0 \leq J \leq 1$ and shown the outcomes in the figure $\ref{f1}$. In the figure, the solid red line indicates the negativity (N) and the dotted blue line represents the CCNR criterion of the system, which will be followed throughout this article.

The figure shows that at the initial condition when there is no interaction in the system $(J=0)$, the negativity (N) of the state is zero, but the CCNR criterion exists. This result implies that at the initial condition, the state is bound entangled and no free entanglement exists in the state. But as the Heisenberg interaction introduces into the system, the negativity of the state increases in oscillatory pattern, and free entanglement produces in the state. The CCNR criterion also follows the negativity and increases in the same pattern due to the interaction. It is noted that this oscillatory pattern does not exhibit the exact sinusoidal behavior and the produced free entanglement becomes zero frequently. This behavior indicates that the sustainability of the free entanglement is low. Further, the frequency of this oscillation depends on the value of the coupling constant $(J)$ and as the value of $J$ increases, the frequency of the oscillation increases, which can be seen in the figure $\ref{f1}$.

\subsection*{Case 2: Dynamics under bi-linear bi-quadratic interaction}

\begin{figure*}
\centering
\includegraphics[scale=1.25]{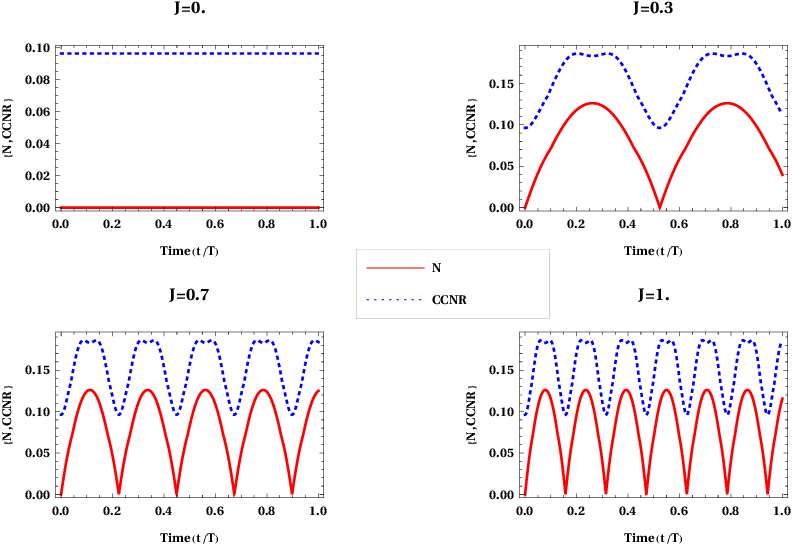} 
\caption{Plot of Negativity(N) and CCNR vs. Time$(t/T)$}\label{f2}
\end{figure*}

The current case deals with the assumption that the interaction between qutrit $A$ and auxiliary qutrit $C$ takes place through a bi-linear bi-quadratic Hamiltonian. The dynamical behavior of the open quantum system under bi-linear bi-quadratic interaction is displayed in the figure $\ref{f2}$ for some selective values of coupling constant $(J)$ in the range $0 \leq J \leq 1$. The figure exhibits that when there is no interaction in the system $(J=0)$, the state has zero negativity (N) and there is no free entanglement in the state. But due to the existence of the CCNR criterion bound entanglement exists in the state similar to the previous case. When the interaction is applied in the system, it almost repeats the case 1 behavior, i.e. negativity and CCNR criterion increases in an oscillatory manner due to the bi-linear bi-quadratic interaction and free entanglement generated in the state. But in this case, the oscillation shows exact sinusoidal behavior and the dropping rate of free entanglement is more frequent than in the previous case, due to the presence of the non-linear term in the Hamiltonian. This phenomenon shows that the sustainability of the originated free entanglement under the bi-linear bi-quadratic interaction is lesser than the Heisenberg interaction. Besides this, the oscillation frequency also follows the previous case and increases with the increment of the coupling constant $(J)$, which is displayed in the figure $\ref{f2}$.

\subsection*{Case 3: Dynamics under DM interaction}

\begin{figure*}
\centering
\includegraphics[scale=1.25]{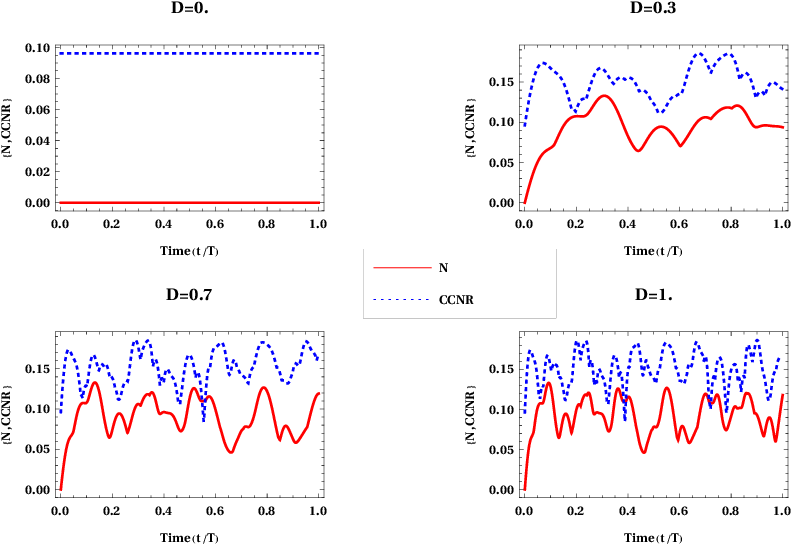} 
\caption{Plot of Negativity(N) and CCNR vs. Time$(t/T)$}\label{f3}
\end{figure*}

This case is framed by considering the DM interaction between qutrit $A$ and auxiliary qutrit $C$. The dynamics of the open system under DM interaction is discussed for the different interaction strength $(D)$ in the range $0 \leq D \leq 1$ and plotted the results in the figure $\ref{f3}$. Analyzing the results it is noticed that for the starting condition $(D=0)$, the system shows a similar result to the previous two cases, i.e. there is no free entanglement in the state, but bound entanglement exists. After introducing DM interaction in the system, it is found that the free entanglement rises in the state due to the interaction. Although the non-sinusoidal behavior of the entanglement is observed in the system, the oscillatory behavior is more distorted than in case 1. But for the current case, the generated free entanglement is sustained for a long time in comparison to the other two interactions.

\section{Conclusion}
In the current article, comparative open quantum system dynamics of the $3 \times 3$ dimensional bipartite bound entangled state, proposed by Bennett \textit{et al.}, have been studied. The study has explored the dynamical analysis under the influence of three different interactions; Heisenberg, bi-linear bi-quadratic, and DM interaction. In the respective cases, the impact of each interaction has been discussed in detail. After analyzing all the cases, it has been found that the Heisenberg interaction and the bi-linear bi-quadratic interaction have produced oscillatory free entanglement for the bound entangled state with the negativity measure. But DM interaction has performed better to sustain the free entanglement over a long period. However, the entanglement pattern is not oscillatory. Here, it is noted that the probability amplitude of the auxiliary state has not played any role in the production of this free entanglement. In short, it can be narrated that the chosen bound entangled state is activated by the introduction of all the interactions mentioned in the article. But in the matter of quality, one can have to be specific since the selected bound entangled state provided by Bennett \textit{et al.} is quite robust against the Heisenberg and bi-linear bi-quadratic interaction. Further, the study can be continued on the mentioned bound entangled state to explore the influence of other interactions.

\end{document}